\documentclass[aps,prl,preprint,groupedaddress]{revtex4}

\usepackage{graphicx}

\begin{document}


\title{Magnetic Determination of $H_{c2}$ Under \linebreak Accurate Alignment in (TMTSF)$_2$ClO$_4$}


\author{J.I. Oh}
\author{M.J. Naughton}
\email[]{naughton@bc.edu}
\affiliation{Department of Physics, Boston College, Chestnut Hill, MA 02467}


\date{Submitted July 1, 2003, Accepted December 9, 2003, Physical Review Letters}

\begin{abstract}
Cantilever magnetometry has been used to measure the upper critical magnetic field $H_{c2}$ of the quasi-one dimensional molecular organic superconductor (TMTSF)$_2$ClO$_4$.  From simultaneous resistivity and torque magnetization experiments conducted under precise field alignment, $H_{c2}$ at low temperature is shown to reach 5T, nearly twice the Pauli paramagnetic limit imposed on spin singlet superconductors.  These results constitute the first thermodynamic evidence for a large $H_{c2}$ in this system and provide support for spin triplet pairing in this unconventional superconductor.
\end{abstract}


\maketitle

The molecular organic conductors (TMTSF)$_2$X are novel electronic materials in the sense that their metallic, insulating and superconducting phases all have unconventional aspects to them.  Due to highly anisotropic structure, strong one and two-dimensional electronic character is observed, such as spin density wave transitions and the quantized Hall effect, respectively, in addition to anisotropic, three-dimensional superconductivity. This latter phase has generated a great deal of interest and its share of controversy since its discovery as the first organic superconductor nearly a quarter century ago [1].  After a rather thorough series of experiments in the early 1980's, the majority opinion concerning the nature of superconductivity in (TMTSF)$_2$X was that they were conventional, albeit anisotropic, BCS superconductors.  Nonetheless, after noting rather effective suppression of superconductivity by nonmagnetic defects [2], Abrikosov suggested in 1983 that there was reason to suspect that quasiparticles paired in a spin triplet state, as opposed to conventional spin singlet [3].  One possible consequence of triplet pairing is the absence of a paramagnetic pair-breaking effect, the ``Pauli limit'' maximum magnetic field in which singlet superconductivity can survive, due to the Zeeman energy difference of oppositely directed spins becoming comparable to the condensation energy [4]. However, resistively-derived critical magnetic field values from early experiments showed little evidence for exceeding the Pauli limit in (TMTSF)$_2$X [5].
   
Lebed [6]  and then others [7] later suggested that orbital pair-breaking, which generally acts independently of the spin effect, could be circumvented in quasi-one dimensional (q1D) superconductors such as (TMTSF)$_2$X by a magnetic field-induced dimensional crossover (FIDC) mechanism for an in-plane aligned field.  In theory, this circumvention could even lead to reentrant superconductivity in very large magnetic fields. Generally speaking, such reentrance would be hard to realize, since spin pair-breaking would kill the superconducting state before FIDC became effective.  That is, unless the q1D system was also a spin triplet superconductor, in which case, FIDC could in principle be tested.  If such a test were to find the dimensional crossover mechanism to be effective, then one would be presented with a situation where {\it no amount of magnetic field could destroy the superconducting state}. Motivated by this idea, a second generation of $H_{c2}$ experiments was initiated, first in (TMTSF)$_2$ClO$_4$ [8]  and then in a sister compound, (TMTSF)$_2$PF$_6$ [9].  In both systems, it was shown that $H_{c2}$ along the in-plane, interchain direction (precisely that predicted by theory to be the most effective for FIDC) significantly exceeded the conventional Pauli limit.  Again, these results were derived solely from electrical resistivity measurements.  In light of the characteristically broad superconducting transition widths (in temperature and magnetic field) observed in (TMTSF)$_2$X materials via resistivity, the exact extent to which $H_{c2}$ exceeds the Pauli limit depends on the resistive criterion one chooses to define $T_c(H)$ or $H_{c2}(T)$.  Using an ``onset'' criterion, $H_{c2}$ was found to exceed $H_P$ by nearly a factor of four in (TMTSF)$_2$PF$_6$ [10], clearly an unconventional situation. On the other hand, if a zero resistance extrapolation were to be used, this value could be rather lower,  potentially making a case for triplet pairing based on $H_{c2}$ less apparent. 

Recently, independent support for triplet superconductivity in (TMTSF)$_2$X was provided by NMR Knight shift ($K_s$)  and tunneling experiments. $K_s$, being a measure of the spin susceptibility, should fall toward zero below $T_c$ for a superconductor with Cooper pairs of zero net spin ($\uparrow\downarrow$), but was instead found for X=PF$_6$ not to change upon entering the superconducting state, as would be expected for a triplet state with equal spins ($\uparrow\uparrow$ or $\downarrow\downarrow$) [11].  Bicrystal junction measurements on X=ClO$_4$ revealed the existence of a large zero bias conductance peak indicative of an Andreev midgap state, interpreted as representing $p$-wave symmetry [12].  On the other hand, another possible mechanism for achieving large critical fields in these materials, distinct from the FIDC model, was recently introduced by Lee, {\it et al.}, involving the formation of slabs of superconductor sandwiched between insulating regions [13].  Furthermore, a recent prediction by Shimahara has a low-field singlet state evolving into a triplet state at high fields [14], mediated by antiferromagnetic fluctuations. Thus, core issues relating to spin (pairing symmetry, and whether these materials are spin singlet Pauli limited or spin triplet unlimited) and orbital  (FIDC versus slabs) angular momenta, each potentially  contributing to a large critical field, require resolution.  A thermodynamic determination of $H_{c2}$ would cement the existence of this as-yet only resistively-determined large critical field.  We provide such a determination in this work.  From  simultaneous torque magnetization and electrical resistivity measurements under accurately aligned magnetic field, we have mapped $H_{c2}(T)$ down to 25  mK, or $T_c$/60. From both measurements, we obtain a zero temperature extrapolation of $H_{c2}$(0) $\cong$ 5T, approximately twice the Pauli limiting field and three times a theoretical limit [15] which accounts for orbital as well as spin effects.

A 0.9$\times$0.4$\times$0.4 mm$^3$   (TMTSF)$_2$ClO$_4$ crystal wired for interlayer resistivity $\rho_{zz}$ measurements.  It was mounted onto a MEMS  magnetometer [16], with gold wires connecting to integrated gold electrodes facilitating simultaneous and independent resistivity and magnetization measurements.  The sample was then mounted onto a stage rotatable about a horizontal axis ($\theta$-rotation) inside a dilution refrigerator, itself attached to a goniometer to provide rotation about the vertical ($\phi$-rotation)  mated to a 13.5T split-coil magnet.  Both rotators provided angular resolution of 0.0025$^\circ$. The resulting $H$-$T$-$\theta$-$\phi$ configuration allowed us to accurately align the sample in any orientation.  

The magnetic signal was calibrated using integrated planar coils on the cantilever through which a current produces a calibrating torque.  The resulting cantilever deflection is detected capacitively with a 1 Hz bandwidth.  The sample was slowly cooled ($\sim$1K/hour), its high quality  quantified by a residual resistivity ratio 
$RRR = \rho_{zz}(300K)/\rho_{zz}(T_o) = 1400 (450)$, where $T_o$ $\sim$ 0K (4.2K).  All the magnetic data were obtained on the second thermal cooldown of the sample.  The $\rho(T)$ curve, RRR value, and resistive $H_{c2}(T)$ phase diagram of this 2$^{nd}$ run were identical to those of the 1$^{st}$, except for the latter below $\sim$0.3K, where a dramatic difference was observed, as  discussed below. 
 
We show in Figure 1 the simultaneously measured torque and resistivity signals at our lowest temperature, 25 mK, for magnetic field precisely aligned along the sample $b^\prime$ direction.  This is the direction within the highly conducting $a$-$b$ layers that is perpendicular to the most conducting chain $a$-axis.  That is, in this triclinic crystal, $b^\prime$ is the projection of the real space lattice direction $b$ onto the plane normal to $a$.  The field is oriented in this direction because theoretically, this is the most favorable direction for FIDC, and empirically, this is indeed where anomalously large critical fields have been observed in transport measurements [8-10].  The inherent ambiguity in defining $H_{c2}$ from transport is evident in Fig. 1: resistivity becomes measurable above about 2.5T, signaling the beginning of the transition out of the superconducting state, with the transition appearing to be complete near 5T (see dashed line extending from the high field, normal state, as well as vertical arrow).  Where one places $H_{c2}$ on such a curve is non-obvious, without the benefit of other physical evidence (note that this problem is quite a bit more severe in the high $T_c$ cuprates).  This evidence is provided by the torque signal. 

First, in referring the torque data in Fig. 1, a hysteretic (irreversible) regime is evident below $\sim$1.3T.  This is but one aspect of a complex superconducting vortex phase uncovered in this material in the process of measuring $H_{c2}$.  Beyond this field, the torque and magnetization are reversible in field sweep direction (i.e. a vortex liquid state), evolving to a well-behaved, $T$-independent, quadratic torque signal at high field in the normal metal phase.  Such a signal in the normal state is consistent with expectations for a clean metal, since both Pauli paramagnetic and Landau diamagnetic susceptibilities are generally $T$- and $B$-independent.   A fit (using data above 7T) to this normal state signal is plotted atop the raw torque signal as (+) symbols and also as a dotted line in the vicinity of 5T.  It is near this field that the measured signal begins to deviate from the normal state background, and we interpret this deviation as a magnetic signature of the upper critical field $H_{c2}$.  Note that even in the conventional description of a type II superconductor-to-normal metal transition in a magnetic field, $H_{c2}$ is a subtle magnetic feature. As the point where the magnetic field, in the form of overlapping superconducting vortices of growing number (in $H$) and size ($\lambda \rightarrow \infty$  as $T \rightarrow T_c(H))$ fully penetrates the sample, thus degrading the diamagnetic susceptibility associated with Meissner currents, $H_{c2}$ is generally marked by only a gradual change in magnetic moment versus field.  Nonetheless, a slope change and departure from the normal state behavior can clearly be seen in the present data, starting at the field indicated by the double arrow.  Note that this field position coincides with the ``onset'' of the resistive transition into the superconducting state, as indicated in the figure.  This validates the use of the transport onset criterion employed in prior reports of $H_{c2}(T)$ in (TMTSF)$_2$X superconductors for $H\parallel b^\prime$.  

A brief discussion about the origin of the magnetic torque, which of course results from a magnetization vector tilted with respect to the applied field, is required.  The symmetry axis for the normal state moment is $b^*$ [17], the normal to the $a$-$c$ planes, and this direction is $\sim 5.5^\circ$ away from the field direction, $H\parallel b^\prime$. This explains the nonzero background torque signal, which we have verified vanishes as $H$ approaches $b^*$.  For the superconducting state, the symmetry axis is $b^\prime$, where, in the absence of shape anisotropy or vortex pinning, there should be no torque signal. The fact that there is a finite signal at $b^\prime$, including in the reversible vortex liquid regime ($\sim$1.3T to 5T in Fig. 1), tells us that both of these terms are present.  We have verified from tilted field studies that the low field torque signal indeed varies as $sin(\theta - \theta_{b^\prime})$, in addition to the small yet finite (and auspicious) shape anisotropy contribution.  

Torque and resistivity data such as those in Figure 1 were taken at several temperatures up to 2K.  We show in Figure 2 representative magnetic moment data derived from the torque, after subtracting the background term discussed above, yielding $\Delta m = \Delta\tau / \mu_o H$, where $\Delta\tau$ is the raw torque less background.  The data shown include both up- and down-sweeps, showing the reversible nature of the magnetization.  On this scale, the onset in decreasing field (above the noise baseline of  $\sim 10^{-12}$ Am$^2$ ($10^{-9}$ emu)) of the diamagnetic signal associated with the formation of the superconducting state becomes evident, as indicated in the figure. At the lowest temperature at which we have torque data, 25 mK, the onset $H_{c2}$ is 4.92T $\pm$ 0.05T.  The resistivity signal collected during that field sweep yielded an onset $H_{c2}$ of 5.02 $\pm$ 0.15T, the larger uncertainty due to the rounded transition characteristic of such measurements (see Fig. 1).  

The resulting phase diagram appears in Figure 3, where we plot onset datum points from magnetization and resistivity field sweeps and from resistivity temperature sweeps.  There are several features of note in this phase diagram. First, the two resistive determinations are well matched by the magnetic $H_{c2}(T)$ over the entire temperature regime.  Second, the zero temperature critical field reaches 5T, close to twice the Pauli limit for singlet superconductivity, defined as $\mu_o H_P = \Delta_o / \sqrt{2}\mu_B = 1.84 T_c=  2.6$T for our sample ($\Delta_o$ is the $T=0$ superconducting energy gap, and $\mu_B$ the Bohr magneton).  In fact, the measured $H_{c2}$(0) is nearly three times a calculated  critical field $H^{LOFF}_{P}$=1.7 T that accounts for both spin and orbital pair-breaking in singlet q1D superconductors [15] including the possibility of an inhomogeneous LOFF state [18].  Third, after a regime of Landau-Ginzburg negative curvature (0.4K$\le T\le$1K), $H_{c2}(T)$ displays positive curvature down to the lowest temperature.  Finally, the inset to Fig. 3 shows a portion of these $H_{c2}(T)$ data replotted along with resistive data from the initial, `virgin' cool of the same sample.  These latter data appear nearly identical to those reported earlier for (TMTSF)$_2$ClO$_4$ [8-10] for this field orientation, with a distinct up-turn in $H_{c2}(T)$ below $\sim$0.25K, and indeed are similar to that reported for (TMTSF)$_2$PF$_6$ [9,10], also for $H\parallel b^\prime$. The overall behavior is very much consistent with that anticipated by the Lebed FIDC effect [6,7,15]: positive curvature developing in $H_{c2}(T)$ as H increases (followed ultimately by reentrant superconductivity at very high fields - as yet not confirmed experimentally).  

We do not have magnetization data to report for this initial cool, but we speculate on the origin of the intriguing difference in $H_{c2}(T)$ below 0.25K between the first and second cooldowns.  As mentioned above, the $\rho(T)$ curves in zero field were indistinguishable between the two runs.  However, the normal state magnetoresistance in the 1$^{st}$ run was significantly larger than in the 2$^{nd}$ ($\sim$20 times so at 25mK).  We suggest that sample microcracks, known to arise in these materials upon cooling, have created interlayer charge channels (i.e. more during the second cooldown) in parallel with the sample's intrinsic interlayer conductance. A simple model mimics the fact that the zero-field $R(T)$ curves for the two successive cooldowns are identical, while those for magnetoresistance $R(H)$ are quite different.  Basically, these microcrack channels short out the intrinsic $H^2$ magnetoresistance at low temperature, when $R_i\gg R_e$, since $R_{zz}(H)=R_e R_i/(R_e+R_i) \sim R_e$ at high field, where $R_i$ is the pristine, intrinsic sample resistance having quadratic magnetoresistance, and $R_e$ is the `extrinsic' field-independent contribution due to microcracks.  The presence of such extrinsic interlayer conduction paths will act to  hinder the ability of a strong magnetic field to decouple the layers, thereby suppressing the FIDC mechanism's ability to facilitate an increase in $H_{c2}$ at high fields and low temperatures. In a pristine / low microcrack density sample, on the other hand, interlayer transport in magnetic fields is dominated by the intrinsic resistivity since, with fewer microcracks, $R_e\gg R_i$, so that $R_{zz}(H) \sim R_i$.  This can explain why the dramatic upturn in $H_{c2}(T)$ seen in the inset for the virgin cool is not as prevalent in the subsequent cool data.  The model also may be used to explain inconsistencies in reported transverse magnetoresistance magnitudes in (TMTSF)$_2$X conductors: each cooldown of each sample has a different microcrack profile. 
 
It is well-established that (TMTSF)$_2$X crystals are easily mechanically ``kinked'' about a (210) dislocation plane [19], causing large jumps in the in-plane resistance, with basically no impact on $T_c$ or $\rho_{zz}$.  It may be that microcracks result from stress-induced kinks of this sort. This relationship was also alluded to by Ishiguro, {\it et al.} [20].  Thus, microcracks should not be considered as impurities in the usual sense, but rather as mesoscopic mechanical deformations that affect the connectivity of the sample, and thus its conductivity.   As the dislocation plane is parallel to the interlayer $c$ direction, the above model can explain the  minimal influence of microcracks on $R_{zz}(T)$, since $R_{e} \gg R_{i}$ in zero field, such that $R_{zz}(T)\sim R_i(T)$ ({\it i.e.} intrinsic).  A future thorough test of this model will require quantifying microcracks and correlating them with $H_{c2}(0)$, with their diminishment possibly facilitating the full impact of FIDC: reentrant superconductivity. 
 
The persistently large critical field observed in this material, now verified from a thermodynamic probe, is not easy to explain in the context of singlet superconductivity.  This fact alone leads us to suggest that the superconductivity is spin triplet in nature.  In conjunction with the complementary experiments mentioned above [2,11,12], this case is now considerably strengthened.

\begin{acknowledgments}
We acknowledge A.G. Lebed and P.M. Chaikin for beneficial discussions, H.I. Ha and J. Moser for assistance, and the support of NSF grants DMR-0076331 and DMR-0308973.
\end{acknowledgments}


\clearpage

\begin{flushleft} 
FIGURE CAPTIONS \\ 
\end{flushleft}

\begin{flushleft} 
{\bf Figure 1} Resistivity (left scale) and torque magnetization (right) in (TMTSF)$_2$ClO$_4$ at 25mK, $H\parallel b^\prime$. The dotted line and + symbols on the torque curve represent a temperature-independent normal state contribution. The onsets of diamagnetism and decreasing resistivity, upon decreasing field, are indicated by the arrow near $H_{c2}\sim$ 5T.  Arrows in the low field vortex state indicate field sweep directions.
\linebreak
\end{flushleft}

\begin{flushleft} 
{\bf Figure 2} Contribution to the magnetization due to superconductivity for $H\parallel b^\prime$ in (TMTSF)$_2$ClO$_4$ at several temperatures.  $H_{c2}(T)$ is obtained at the onset of finite moment $\Delta m(H)$, as indicated for $T$=25mK.
\linebreak
\end{flushleft}

\begin{flushleft} 
{\bf Figure 3} Upper critical field $H_{c2}$ along the $b^\prime$-axis in  (TMTSF)$_2$ClO$_4$, from both resistivity and magnetization. Inset also shows resistive $H_{c2}$ data from the same sample's initial cooldown.
\end{flushleft}
\clearpage
\begin{figure}
\includegraphics[width=7in]{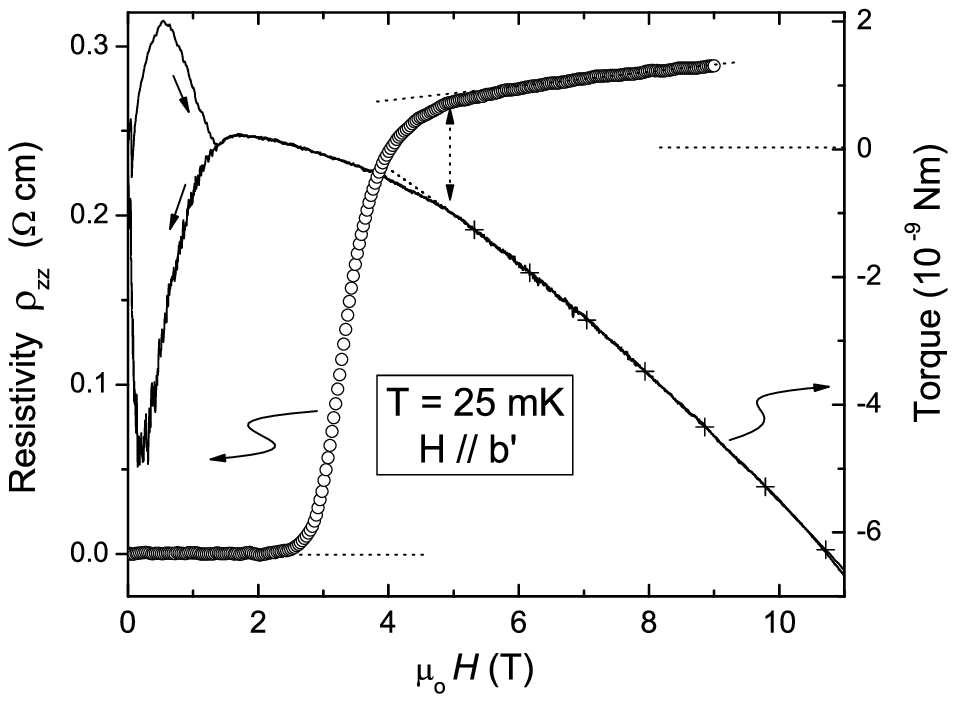}
\end{figure}
\bigskip
Figure 1 --- Oh \& Naughton, 2003
\clearpage
\begin{figure}
\includegraphics[width=7in]{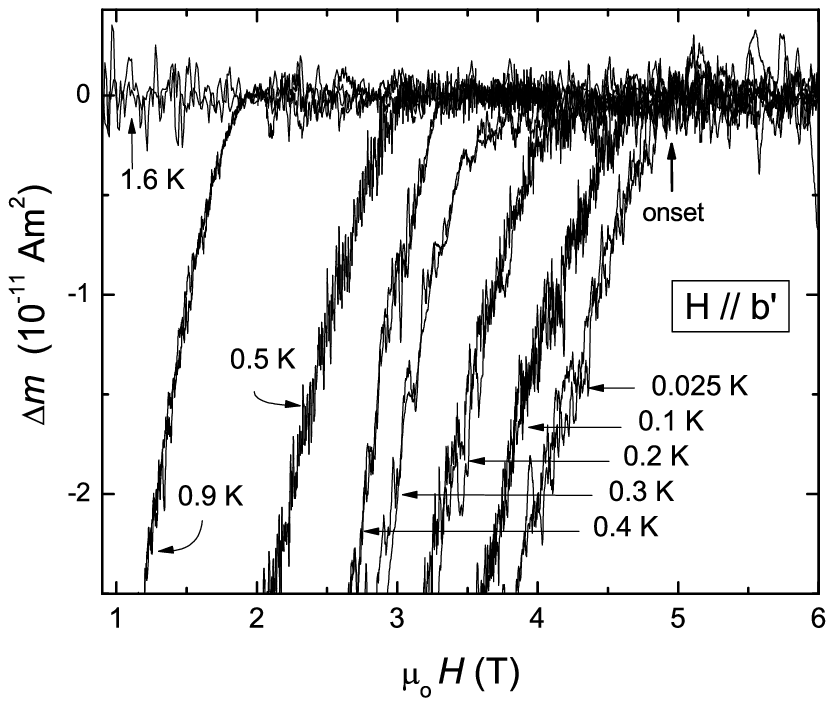}
\end{figure}
\bigskip
Figure 2 --- Oh \& Naughton, 2003
\clearpage
\begin{figure}
\includegraphics[width=7.5in]{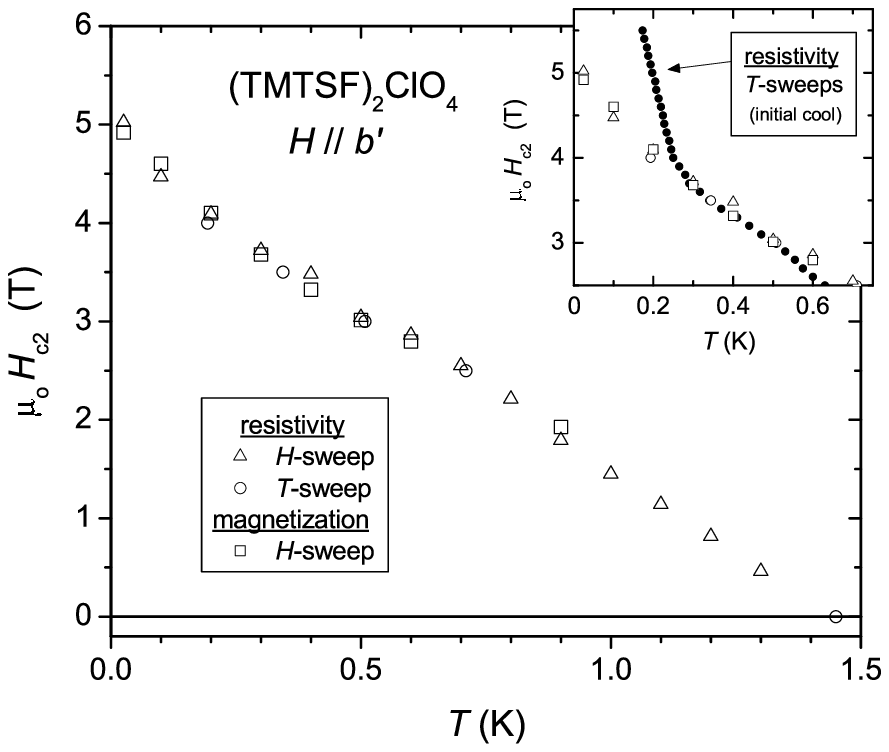}
\end{figure}
\bigskip
Figure 3 --- Oh \& Naughton, 2003

\end{document}